\RequirePackage[2020/08/10]{latexrelease}
\documentclass[aps,pra,reprint,superscriptaddress,longbibliography]{revtex4-2}

\usepackage{physics}
\usepackage{amsmath}
\usepackage{amssymb}
\usepackage{graphicx}
\usepackage{float}
\usepackage{mathrsfs}
\usepackage{mathtools}
\usepackage{extarrows}
\usepackage[margin=0.7in]{geometry}
\graphicspath{{PNG/}}
\usepackage{xcolor}

\begin{document}
	
	\title{Edge state behavior of interacting Bosons in a Su-Schrieffer-Heeger lattice}
	
	\author{A.~ Ghosh}
	
	\affiliation{School of Physics, University of Melbourne, Melbourne, 3010, Australia}
	
	\affiliation{Indian Institute of Technology Kharagpur, Kharagpur 721302, West Bengal, India}

	\author{A.~M.~Martin}
	
	\affiliation{School of Physics, University of Melbourne, Melbourne, 3010, Australia}

	\date{\today}

	\begin{abstract}
	In the low momentum regime, the Su-Schrieffer-Heeger (SSH) model's key characteristics are encapsulated by a Dirac-type Hamiltonian in continuum space, i.e., the localized states emerge at the boundaries. Building on this, we have developed an effective Hamiltonian to model ultra cold interacting Bosons on an SSH like lattice through variational minimization under the mean field approximation. To pinpoint the boundary states, we have developed an algorithm by generalizing the imaginary time propagator, where a initial state evolves under the squared Hamiltonian to converge to the targeted state. This algorithm has broader applicability, enabling the identification of specific eigenstates in various contexts. Furthermore, we draw a parallel to an experimentally physical setup involving a gas of ultra cold Bosons confined to an array of potential wells with alternating depths. By establishing the system's analogy with the SSH system, we apply our algorithm to investigate boundary states in the presence of interaction, demonstrating how these findings align with those of the continuous system.
		
	\end{abstract}
	
	\maketitle

	\section{Introduction \label{sec:1}}
Topological physics represents a dynamic frontier within condensed matter physics \cite{hasan2010colloquium, rachel2018interacting, goldman2016topological, dmytruk2022controlling, ghosh2023quench}. In these systems, there is a topology associated with the bulk depending upon the symmetries present, giving rise to exotic edge/boundary properties, and this is referred to as bulk boundary correspondence \cite{asboth2016short, shen2012topological, bernevig2013topological}. The properties of edge states are critically significant and can manifest in diverse contexts, such as in the Su-Schrieffer-Heeger(SSH) model, the quantum Hall effect, the quantum spin Hall effect, in higher-dimensional topological insulators etc. \cite{kane2005quantum, kane2005z, benalcazar2017quantized, schindler2018higher, asboth2016short, bernevig2013topological, shen2012topological}. The experimental achievement of various topological phases has generated considerable excitement and holds great promise \cite{jotzu2014experimental, leonard2023realization, osterwalder2009observation, konig2007quantum, de2019observation}. Numerous techniques have been employed to probe the topological states, advancing our understanding and paving the way for innovative applications \cite{yin2021probing, peng2017observation, zhang2018observation}. They are implemented through a variety of platforms, including optical lattices, where ultracold atoms are manipulated in controlled environments, photonic systems that guide light through engineered structures, and diverse solid-state materials that provide unique opportunities to explore and exploit topological phenomena \cite{nicolau2023bosonic, meier2016observation, lu2014topological}. Additionally, topological phases are being studied in systems outside the quantum regime, such as electrical systems and metamaterials \cite{paulose2015topological, lee2018topolectrical}. 
The Su-Schrieffer-Heeger (SSH) model \cite{su1979solitons, asboth2016short, shen2012topological}, a foundational framework for hosting topological phases, has successfully been implemented in experimental settings \cite{nicolau2023bosonic, meier2016observation, de2019observation}. 
These studies naturally prompt the fascinating question of how boundary states in the SSH system, primarily determined by a single-particle Hamiltonian, transform with the introduction of interactions. Several recent studies have investigated this phenomenon across various cases and settings \cite{mikhail2022quasiparticle, grusdt2013topological, tuloup2020nonlinearity}. For example, \cite{mikhail2022quasiparticle} examined the impact of interactions on topological edge states in spinful Fermions, specifically examining the Su-Schrieffer-Heeger-Hubbard (SSHH) model. Similarly, \cite{grusdt2013topological} explored this scenario in the context of the Bose-Hubbard model and \cite{tuloup2020nonlinearity} examined the persistence of topological characteristics amid competition from chiral symmetry-breaking potentials and on-site interaction terms in a tight-binding setup.\\
Leveraging on the recent breakthroughs in ultra-cold physics that have unlocked the ability to create condensates in non-ground states using cesium atoms \cite{horvath2024bose}, in this paper, we shed light on the fate of the boundary states in the case of dilute interacting Bosons gas in ultra-cold settings on an SSH-like lattice. The emphasis is on boundary states, investigated using a constructed effective Hamiltonian, which we analyze using a developed algorithm.  \\
The SSH lattice under low momentum is described by a continuum Dirac-type Hamiltonian \cite{takayama1980continuum, shen2012topological}, which captures the highly localized states at the boundary. Taking this into account, in Section II, we employ a mean field approximation and perform variational minimization of the energy functional to obtain an effective Hamiltonian for the system, assuming contact interactions between particles corresponding to the s wave scattering. This Hamiltonian, being a derivative of a Dirac-type Hamiltonian, can host a spectrum of eigenvalues, both positive and negative. 
As the conventional imaginary time algorithm probes for the lowest eigenvalue \cite{dalfovo1996bosons}, in section III A, we develop an algorithm to pinpoint a particular eigenstate in order to probe the boundary state. In subsection III B, we prepare the ground for the application of this algorithm to this particular problem. In Section IV A, we explore boundary states across a range of interaction strengths and boundary sharpness levels, comparing these findings to the situation where no interactions are present. To cultivate an intuitive and physically meaningful comprehension of the results, subsection IV B investigates a related scenario involving bosons in a periodic well-like lattice with staggered amplitudes. After demonstrating its analogy to the SSH system, we proceed to apply the developed algorithm to investigate the boundary eigenstates in the presence of interactions. These eigenstates are analyzed using the Gross-Pitaevskii equation, allowing us to draw insightful parallels between the two systems. Finally, in Section V, we provide a comprehensive summary and conclusion of our findings.
	
	\section{Exploring Continuum Mechanics in a SSH system with Interactions \label{sec:2}}
A discrete tight binding system such as the SSH system can be treated as a continuous system under an appropriate regime., i.e., under low momentum approximation\cite{asboth2016short, takayama1980continuum}. This continuum model also explains the presence and localization of edge states at the boundary between two distinct topological regimes. The central idea here is incorporating interactions to develop an effective model for describing the Bosonic interacting SSH system, enabling the study of edge states. To begin with, we provide a brief overview of the continuum treatment of the SSH system. As a starting point, the SSH Hamiltonian ($  \hat{H} $) in discreet space is given by,
\begin{equation}\label{key}
	\hat{H} = v\sum_{m=1}^{m=N}\ket{m,A}\bra{m,B} + w\sum_{m=1}^{m=N-1}\ket{m,A}\bra{m+1,B}+h.c ,
\end{equation}
where, $ m $ denotes the index of the unit cell, which is composed of sublattices $ A $ and $ B $, linked by staggered hopping amplitudes. The parameter $ v $ corresponds to the intra-cellular hopping amplitudes, while $ w $ represents the inter-cellular hopping amplitudes.
Translational invariance allows us to write the Hamiltonian in momentum space using $ \ket{K}=\ket{k,\alpha} $ for $ \alpha \in \{A,B\} $, 
\begin{equation}\label{key}
	\hat{H}(k)= [v+w\cos k]\hat{\sigma}_{x} + w\sin k\hat{\sigma}_{y}.
\end{equation}
Considering points in the momentum space close to the point of closure of the band $(q = k-k_{0})$,
we have
\begin{equation}\label{key}
	\hat{H}(q+k_{0}) = [ v+w\cos(q+k_{0})] \sigma_{x} + w\sin(q+k_{0})\sigma_{y}.
\end{equation}
When $ q \approx 0 $ and $ k_{0} = \pi  $, 
\begin{equation}\label{key}
	\hat{H}(q) = [v-w]\hat{\sigma}_{x}-wq\hat{\sigma}_{y}.
\end{equation}
In continuum treatment of a discrete system, the  $ q $ is replaced by the corresponding operator $ - i \hbar \partial/\partial x  $, treating the space as continuos so that, $ \ket{\psi_{Discrete}} = \sum_{m}\phi(x=m)\ket{m} $, if $ \phi(x)  $ is the solution in continuum space. This is known as the envelope function approximation \cite{asboth2016short}. For this to hold, the wavefunction in momentum space must be confined well within the boundaries of the first Brillouin zone.
Here, we get a Dirac type equation (4) with a mass term $ M = v-w $. For the case when $ v = w $, the mass term vanishes, and we have a linear dispersion relation. The boundary between systems of different topological constants hosts an edge state, and this fact is also reflected through equation 4, i.e., when we have a change in sign in $ M $ ($ v $ goes from less than $ w $ to greater than $ w $), theoretical calculations show there is an edge state on the boundary with exponential decay\cite{asboth2016short, takayama1980continuum}. Since we intend to examine the edge state under interactions, we will address this through numerical methods, as we will shortly see in Section III. \\
Now under the assumptions of this model, if there are $N_{p}$ interacting Bosons, the total Hamiltonian will be,
\begin{equation}\label{key}
	\hat{H} = \sum_{i=1}^{i=N_{p}}M_{i}\sigma_{x}-wq_{i}\sigma_{y}+\frac{1}{2}\sum_{i}\sum_{j}V(x_{i}-x_{j})*I
\end{equation}
In order to get the effective governing equation for the system, we treat the Hamiltonian in a mean field manner by considering the ansatz wavefunction, 
\begin{equation}\label{key}
	\Psi= \psi_{1}\otimes\psi_{2}\otimes\ldots\otimes \psi_{N_{p}-1}\otimes\psi_{N_{p}}
\end{equation}
here, \( \psi_n \) represents the wavefunction of the \( n^{th} \) particle and is expressed as a \( 2 \times 1 \) matrix as a function of \( x_{n} \), where the two components correspond to the two sublattices. Minimizing the functional $ \bra{\Psi}\hat{H}\ket{\Psi} $ respecting the constraint $ \bra{\Psi}\ket{\Psi}=1 $,	which in coordinate space is,
\begin{align}\label{key}
	&\int\psi_{1}^{*}\otimes\ldots\otimes\psi_{N_{p}}^{*}[\sum_{i=1}^{i=N_{p}}M_{i}\sigma_{x}-wq_{i}\sigma_{y} +\\\nonumber &\frac{1}{2}\sum_{i}\sum_{j}V(r_{i}-r_{j})*I]\psi_{1}\otimes\ldots\otimes\psi_{N_{p}}dx_{1}..dx_{N_{p}}
\end{align}
For ultra cold bosons, all particles acquire the state $ \psi $, so the first term in the equation simplifies to $ N_{p}\int \psi^{\dagger} M\hat{\sigma}_{x} \psi dx $. The second term, after applying integration by parts, reduces to $ -N_{p} w\int i\hbar\frac{d\psi^{\dagger}}{dx}\hat{\sigma}_{y}\psi $. For dilute gases at low temperatures, the interaction can be described by \(V(x) = \alpha\delta(x)\). Thus, the third term in the integral is given by,$ \frac{N_{p}(N_{p}-1)}{2}\int  \psi^{*}_{i}\psi^{*}_{j}\alpha\delta(x_{i}-x_{j})\psi_{i}\psi_{j}dx_{i}dx_{j} $. 	
So the minimum of the functional in equation 7 translates to,
\begin{align}\label{key}
	&\delta\int[N_{p} \psi^{\dagger} M\hat{\sigma}_{x} \psi dx - N_{p} w i \hbar \frac{d\psi^{\dagger}}{dx}\hat{\sigma}_{y}\psi +\nonumber\\ &\frac{N_{p}^{2}}{2}\alpha\psi^{*}\psi^{*}I\psi\psi-EN_{p}\psi^{*}\psi]dx=0
\end{align}
	Applying Euler-Lagrange equation, $ \partial J / \partial \psi^{*}-\partial/\partial x(\partial J/\partial \psi^{*}_{x})= 0 $, we have the mean field equation for the interacting bosons on an SSH lattice,
	\begin{equation}\label{key}
		\left[ M\sigma_{x}-w\hat{q}\sigma_{y}+\alpha^{\prime}|\psi|^{2}I  \right]\psi = E\psi
	\end{equation}
	In the above discussion, \( J \) is the functional in Equation 8, and \(\alpha^{\prime}\) is \(N_{p}\alpha/2\).
	
	This equation—a massive Dirac-type equation enhanced by a nonlinear term representing interacting Bosons on an SSH lattice—will enable us to explore the edge states, as will be discussed in Section IV.
	As the Dirac equation has infinitely many negative eigenvalues possible, it is difficult to obtain the eigenstates corresponding to the near zero energy eigenstates by using the imaginary time propagation algorithm alone. To tackle this problem, we have devised an algorithm to compute a chosen eigenvector, which is particularly useful when it is not in the ground state. Furthermore, we have applied the algorithm to solve a physically intuitive continuum system with similar properties, as is discussed in Section IV(B). The next Section(III) is dedicated to discuss the algorithm.
	\section{Introduction to the algorithm of imaginary time propagator under squared Hamiltonian}
	\subsection{The algorithm}
	The focus here is on the localized states at the boundary and the impact of interaction on them; however, equation $ 9 $ hosts number of states with both negative and positive eigenvalues. The widely used imaginary time propagator algorithm is designed to provide the eigenvector corresponding to the lowest eigenvalue \cite{dalfovo1996bosons}. For the sake of completeness, the basic principle of operation may be stated as follows: For a system, $ \hat{H}\psi=E\psi $, where $ E $ can be $E_{1}<E_{2}<E_{3}\ldots$. A trial wavefunction is evolved according to $ \psi(t)=\sum_{n}c_{n}e^{-it E_{n}}\psi_{n} $, where $ c_{n} $ is the overlap between the trial state and the $ n^{th} $ true state of the system. If we consider the imaginary time, i.e., $ \tau = it $, then $ \psi(t)=\sum_{n}c_{n}e^{-\tau E_{n}}\psi_{n} $. Now as $ \tau $ increases, the contribution from the terms with higher energy decays faster than the lowest energy term. After a sufficient number of steps, with normalization at each step, the lowest energy state is exponentially enhanced. \\
	The algorithm we devised deals with this shortcoming. The basic idea is to update the wavefunctions using $ \hat{H}^{2} $ corresponding to the square of  energy, i.e., for $ \hat{H}\psi = E_{0}\psi $, updating the trial wavefunction $ \psi_{0} $ at each step using $ (\hat{H}-E_{0})^2 $ to reach eigenvector $ \psi $ corresponding to the lowest eigenvalue close to zero, for close enough trial $E_{0}$ for the desired eigenvector. 
	Consider the increment $ \psi $ with respect to the imaginary time parameter in the direction of the negative gradient of the square of energy in wavefunction space:
	\begin{equation}\label{key}
		\frac{\partial\ket{\psi}}{\partial \tau} = -\frac{1}{2}\frac{\partial E^{2}}{\partial \ket{\psi}}.
	\end{equation}
	Noting $ E = \bra{\psi}\hat{H^{'}}\ket{\psi} = \bra{\psi}\hat{H}-E_{0}\ket{\psi} $, we obtain, $ \frac{\partial \ket{\psi}}{\partial \tau} = \hat{H'}^{2}\ket{\psi} $, leading to the time evolution as
	\begin{equation}\label{key}
		\ket{\psi(\tau+\delta \tau)}\approx \left[\frac{I - \delta \tau * \hat{H'}^{2}}{I + \delta \tau * \hat{H'}^{2}}\right] \ket{\psi(\tau)},
	\end{equation}
 and the convergence is achieved when $ \partial \ket{\psi}/ \partial \tau \approx 0$, i.e., $ E(\tau+ \delta \tau)\approx E(\tau) $. Figure 1 depicts the steps involved in the algorithm of the imaginary time propagator under the squared Hamiltonian (ITPUSH), which is used to tackle the problem of analyzing the boundary state of interacting ultra-cold Bosons on an SSH lattice. The value of $ E_{0} $ and the trial wavefunction obtained by solving the case of no interaction using the finite difference method is fed into equation 11 for iteration until convergence is obtained, which is subjected to verification using the stability of the probability density under real-time evolution of the wavefunction ($ \psi(t)=\sum_{n}c_{n}e^{-it E_{n}}\psi_{n} $).  The interaction strength is gradually increased, using the wavefunction from the previous interaction strength as the trial wavefunction. The following subsection demonstrates the application of the algorithm developed to study the boundary states of interacting Bosons in an SSH lattice.\\
	\begin{figure}
		\centering
		\includegraphics[width=1.0\linewidth]{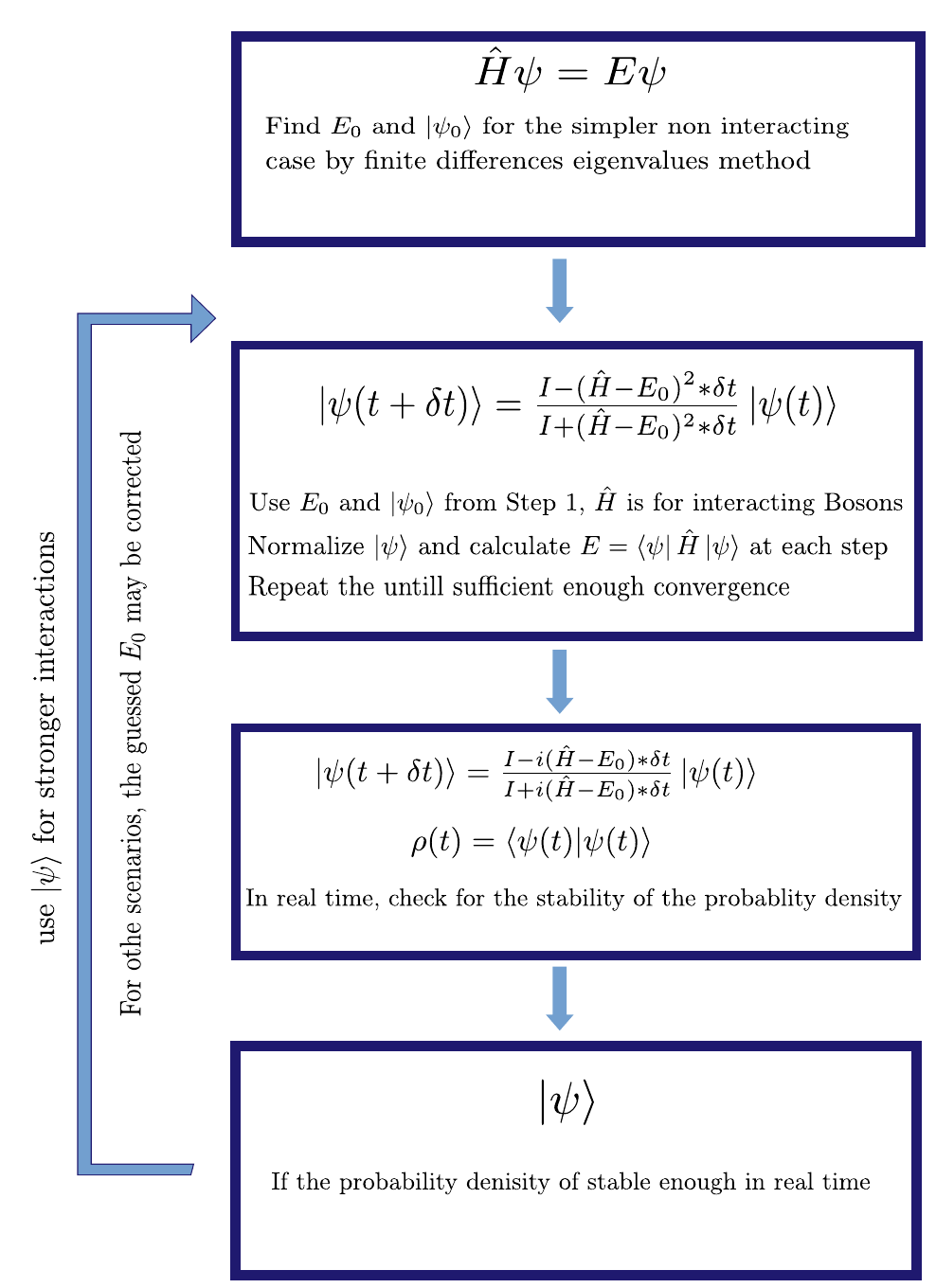}
		\caption[write a suitable caption here]{ Flow chart illustrating the implementation of the algorithm to identify the boundary localized state of interacting bosons in an SSH lattice, whose energy lies embedded within the system's energy spectrum.}
		\label{fig:flowchart}
	\end{figure}
The significance of the algorithm lies in its ability, in principle, to reach any desired target state, provided the trial wavefunction $ \psi_{0} $ and the energy $ E_{0} $ are sufficiently close. The selection of $ E_{0} $ must ensure that its difference from the actual energy is smaller than its difference from any nearby energy levels.

	\subsection{The implementation}
	Our strategy involves setting up the matrix equation \( H\psi = E\psi \) for the system, and then moving on to determine the \( H^2 \) term for use in the ITPUSH algorithm. To begin with, the matrix form of equation 9 is,
	\begin{equation}\label{key}
		    \begin{bmatrix}
			\alpha^{\prime}(|\psi_{a}|^{2}+|\psi_{b}|^{2}) & M+w\frac{\partial}{\partial x}\\
			M-w\frac{\partial}{\partial x} & \alpha^{\prime}(|\psi_{a}|^{2}+|\psi_{b}|^{2})
			    \end{bmatrix}\begin{bmatrix} \psi_{a}\\ \psi_{b}\end{bmatrix}= E\begin{bmatrix} \psi_{a}\\ \psi_{b}\end{bmatrix}
	\end{equation}
	The expression for $ H^{2} $ will then be,
	\begin{widetext}
	\begin{equation}\label{key}
    \begin{aligned}
		\begin{bmatrix}
			\gamma^{2}+M^{2}+w\frac{\partial M}{\partial x}-w^{2}\frac{\partial^{2}}{\partial x^{2}} & 2\gamma M + 2\gamma w \frac{\partial}{\partial x}+ w\frac{\partial \gamma}{\partial x}\\ 2\gamma M - 2\gamma w \frac{\partial}{\partial x}- w\frac{\partial \gamma}{\partial x} & \gamma^{2}+M^{2}-w\frac{\partial M}{\partial x}-w^{2}\frac{\partial^{2}}{\partial x^{2}}
		\end{bmatrix}\begin{bmatrix} \psi_{a}\\\psi_{b} \end{bmatrix}=E^{2}\begin{bmatrix} \psi_{a}\\ \psi_{b} \end{bmatrix},
	\end{aligned}
	\end{equation}
	\end{widetext}
where $ \gamma = \alpha^{\prime}(|\psi_{a}|^{2}+|\psi_{b}|^{2}) $.
This leads us to coupled differential equations in $  \psi_{a}$ and $ \psi_{b} $. To solve for $ \psi = [\psi_{a} \quad \psi_{b}]^{T} $,  under the ITPUSH algorithm  we write $ \psi $ as 
\begin{equation}\label{key}
	\psi_{a}\otimes[1\quad 0 ]^{T} + \psi_{b} \otimes [0\quad 1]^{T}
\end{equation}
  and $  \hat{H} $ as 
\begin{equation}\label{key}
	\hat{H} = \hat{H}_{11}\otimes \begin{bmatrix}1 & 0\\ 0 & 0  \end{bmatrix} + \hat{H}_{12}\otimes\begin{bmatrix} 0 & 1\\0 & 0 \end{bmatrix} + \hat{H}_{21}\otimes \begin{bmatrix} 0 & 0\\1& 0 \end{bmatrix}+ \hat{H}_{22}\otimes \begin{bmatrix} 0& 0\\ 0& 1 \end{bmatrix}
\end{equation}
In the above steps the space is discretized into a grid of $ N_{g} $ points, making $ \psi $ and $ \hat{H} $ correspond to a grid of $ 2N_{g} $ and $ 2N_{g}*2N_{g} $points, repsetively. This enables us to calculate for $ \psi_{a} $ and $ \psi_{b} $ simultaneously with the algorithm. 
As discussed in the previous subsection, after sufficient convergence in $ E $ using equation 11, the obtained $ \psi $ is then verified by evolving in real time for stability of the probability density under the Hamiltonian $ \hat{H} $ (Equation 12 ). We then move on to stronger interaction using the result from the current interaction as the fodder for the algorithm and this cycle is repeated. 
	\section{Result and Disscussions}

	\subsection{Case of Continum SSH system}
		In this section, we will explore the application of the algorithm developed, for the system with a boundary, specifically when there is a sign change in M(x) ($ =v-w  $) along the length of the system(see Figure 2(b).). We start by investigating the case when no interaction is present, i.e.,  $ \alpha^{\prime} = 0 $. The boundary of the domains can have different patterns, which we denote by the parameter $ \beta $, such that $  M(x) $ is given by $ M(x)= -M_{0} tanh(\beta x) $. This quantifies the steepness of the domain wall. 
    \begin{figure}[h]
    	\centering
    	\includegraphics[width=1.0\linewidth]{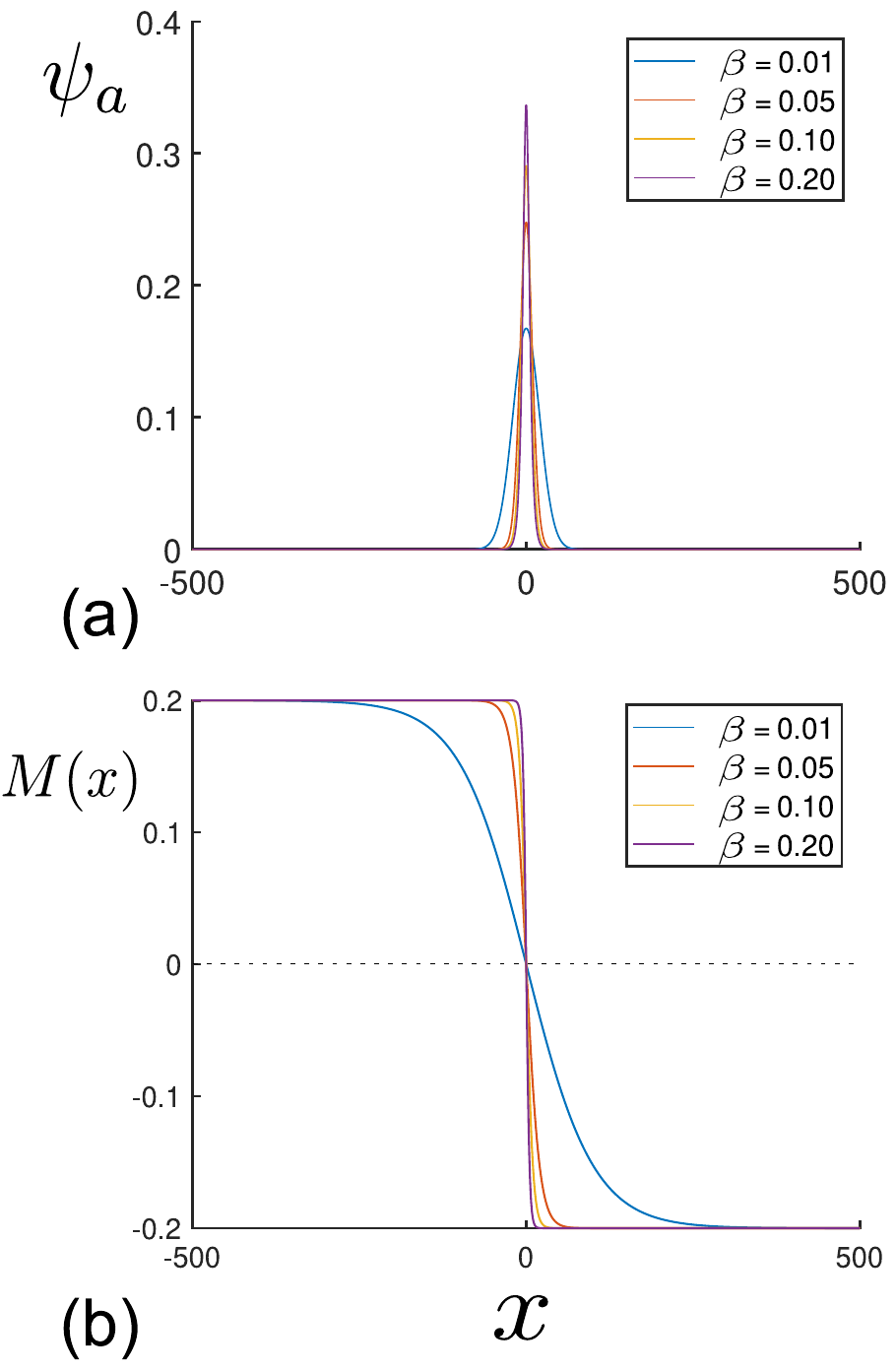}
    	\caption{(a) Displays the variation of the boundary states \( \psi_{a} \) corresponding to sublattice \( a \) as a function of the parameter \( \beta \), which characterizes the steepness of the domain wall shown in (b).}
    	\label{fig:wavefunctiondomain}
    \end{figure}
Drawing from the discrete model (Equation 1) and analytical works on continuum treatments of the SSH system \cite{takayama1980continuum, asboth2016short,shen2012topological}, it is well-understood that the boundary states possess zero energy. Accordingly, we designate \( E_{0} = 0 \) in the ITPUSH algorithm to derive the corresponding wavefunction. This choice aligns with the theoretical predictions and ensures that the algorithm accurately captures the characteristics of the boundary states.
Figure 2(a) provides a vivid visualization of localized states at the domain boundary, underscoring a critical characteristic of boundary states: the extent of their localization is governed by the sharpness of the boundary. Specifically, sharper boundaries result in more localized states, while flattened boundaries lead to greater delocalization. This relationship offers key insights into the behavior of boundary states. Importantly, across all cases examined, the \( \psi_{b} \) consistently remains at zero.
As detailed in Section III, the verification of the converged wavefunction is confirmed by its stability during real-time evolution under \( \hat{H} \). This is exemplified in Figure 3, where the probability density corresponding to $ \psi_{a} $ remains constant throughout the real-time evolution of the wavefunction, demonstrating its robustness and accuracy.
\begin{figure}
	\centering
	\includegraphics[width=1.0\linewidth]{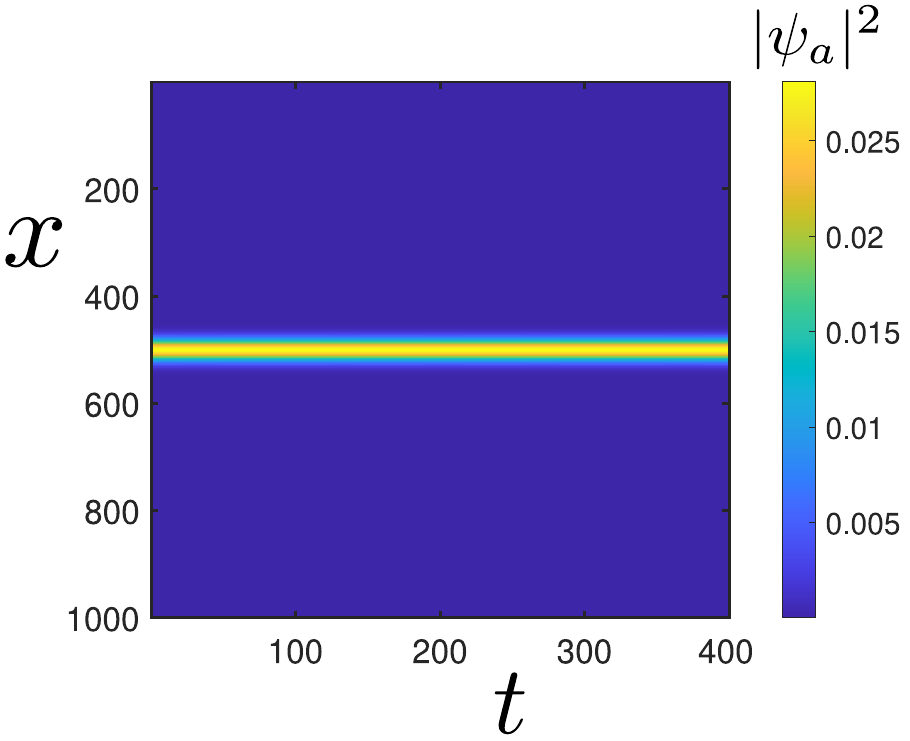}
	\caption{Probability density of the numerically obtained boundary state wave function \( \psi_{a} \) corresponding to sublattice \( a \), plotted as a function of real time \( t \). The wave function is for the case \( \beta = 0.01 \) and \( \alpha' = 0 \). The color scale indicates the magnitude of the probability density, which remains constant during the evolution.}
	\label{fig:convergencecartoon}
\end{figure}
\\
We move on to consider the scenario where interactions are present in the system, with the mean-field governing equation described by equation 9. The developed algorithm is applied once more; in this instance, we initiate the process by utilizing the wavefunctions $\psi_{a}$ and $\psi_{b}$ derived from the non-interacting case as the initial state for our analysis. The impact of the interaction is illustrated by adjusting its strength across two distinct levels of boundary sharpness(Figure 4). This variation allows observation of how the interaction influences the system under different boundary conditions. Unlike the no-interaction case where sub-lattice B remained unpopulated due to the chiral symmetry of the Hamiltonian, the introduction of interaction changes this dynamic. As the interaction strength increases, sub-lattice B becomes progressively more populated, which, in turn, reduces the population of sub-lattice A(refer to Figure 4 ). The SSH nature of the parent Hamiltonian tends to confine the wavefunction to the A sub-lattice. However, increasing the interaction strength counters this tendency. Since a localized wavefunction results in higher potential energy, the system tends to delocalize by occupying sub-lattice B to lower the potential energy. This effect is similar to what is observed in the Gross-Pitaevskii equation, where, with the increase in the interaction, the central maxima of the ground state wave function decreases and flattens in order to compensate for the increased potential energy arising due to repulsive interactions \cite{dalfovo1996bosons}. In the typical Gross Pitaevskii equation, the external trap of harmonic potential is the factor that tries to localize the wavefunction, whereas, in the case of this system, it is the topological barrier. There exists a competition between the interaction and the topological barrier in determining the structure of the wavefunction.
\begin{figure}
	\centering
	\includegraphics[width=1.0\linewidth]{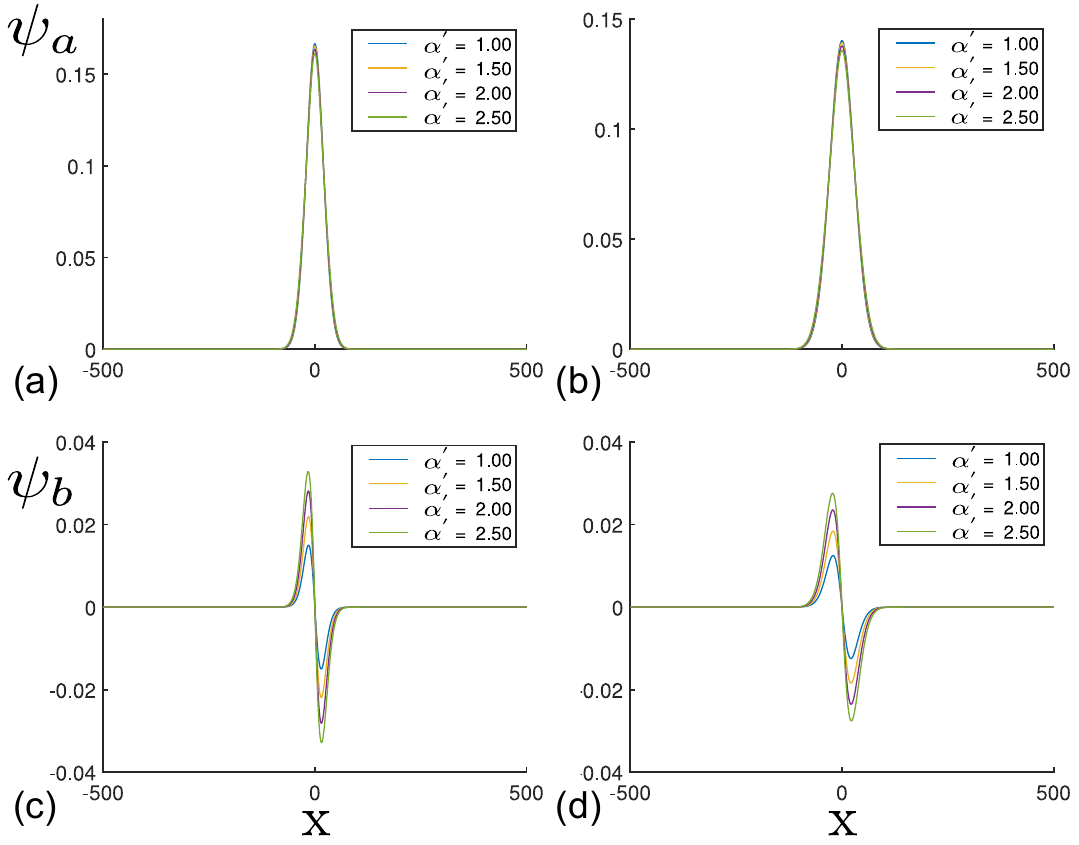}
	\caption{The impact of interaction on boundary states for two levels of boundary stiffness is illustrated. The two columns depict different boundary sharpness values: \( \beta = 0.01 \) for the left column and \( \beta = 0.005 \) for the right column. The rows ((a), (b) and (c), (d)) show the boundary wavefunctions for sublattices \( a \) and \( b \), respectively. With increasing interaction parameter \( \alpha^{'} \), the population in sublattice \( b \) increases.}
	\label{fig:wavefunctioninteraction2}
\end{figure}

	\subsection{Analogous physical system}
	Here, we consider a physically intuitive and implementable system to visualize the results obtained in Subsection A of this Section. The system consists of a ultra-cold Boson gas confined in a periodic potential well with alternating amplitudes, truncated at both ends by a step potential, as illustrated in Figure 5. The behavior of ultracold bosons in a trap is governed by the Gross-Pitaevskii equation, a nonlinear partial differential equation that describes the macroscopic wave function of the condensate. In its dimensionless form, the 1D Gross-Pitaevskii equation is given by,
\begin{equation}
i \frac{\partial \psi}{\partial t} = \left( -\frac{\partial^2}{\partial x^{2}}+ V_{ext}(x) + g |\psi|^2 \right) \psi ,
\end{equation}
with $ \psi $ being the macroscopic wavefunction, $ V_{ext} $ being the externally applied potential, and $ g $ is the interaction strength \cite{dalfovo1996bosons, pethick2008bose}.
	
	\begin{figure}
		\centering
		\includegraphics[width=1.0\linewidth]{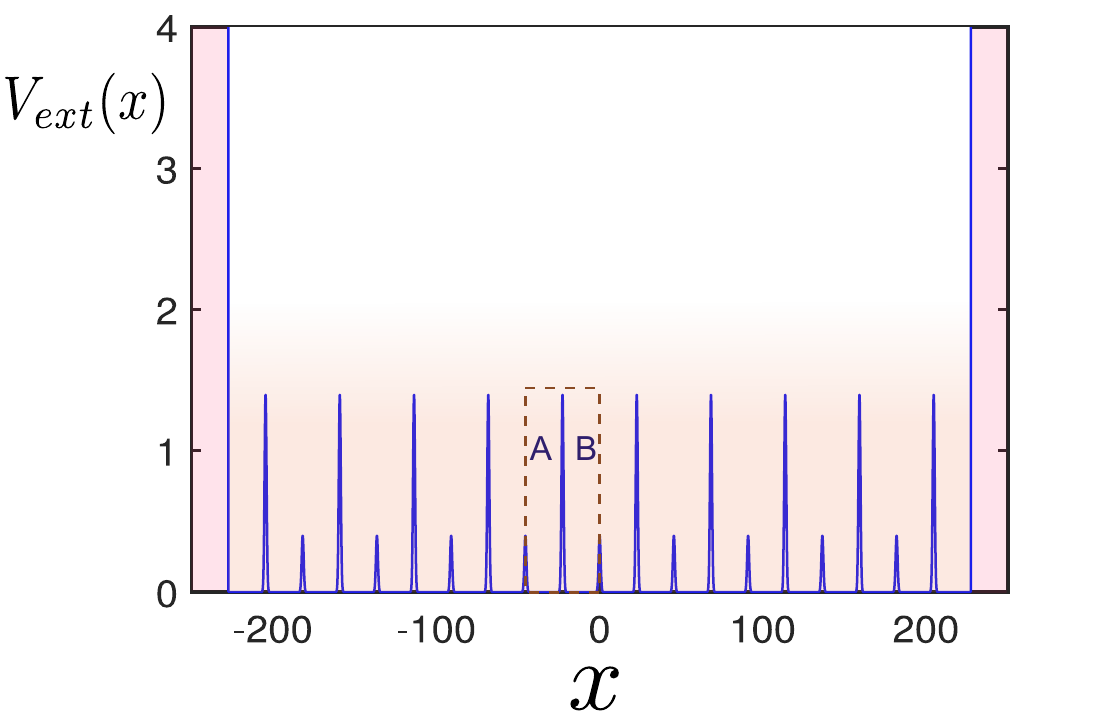}
		\caption{Schematic diagram of the system: Inside the box, sharp Gaussian potentials with alternating amplitudes are applied, while the box itself is characterized by higher step-like potentials. This configuration forms two sublattices (A and B) with staggered tunneling amplitudes between the sites and open boundary conditions at the edges. The gradated cream colour is a cartoon representation of the region containing ultracold bosonic gas, while the light pink stripes at the ends denote the open boundaries.}
		\label{fig:externalpotential}
	\end{figure}
	The periodic structure with a series of equidistant alternating potentials is formed by sharp Gaussians with two corresponding amplitudes, 
	\begin{equation}\label{key}
		V_{1(2)}(x) = C_{1(2)}e^{-d(x-\mu)^{2}},
	\end{equation}
	whereas the ends are modeled by step potentials $ V_{step} $, with a magnitude much higher than that of $ C_{1(2)} $. The total form of potentials may be mathematically stated as
	\begin{equation}\label{key}
		\begin{aligned}
		V_{ext}=&\sum_{n=2}^{N_{pot}-1}[V_{2}[1-mod(n,2)]+V_{1}[mod(n,2)]]\\
		&+V_{step}[\delta_{1,n}+\delta_{N_{g},n}],
		\end{aligned}
	\end{equation}
    where, $ N_{pot} $ is the total number of potentials including those of the boundary, represented by indices $ n=1 $ and $ n=N_{pot} $.
    \begin{figure*}
    	\centering
    	\includegraphics[width=1.0\linewidth]{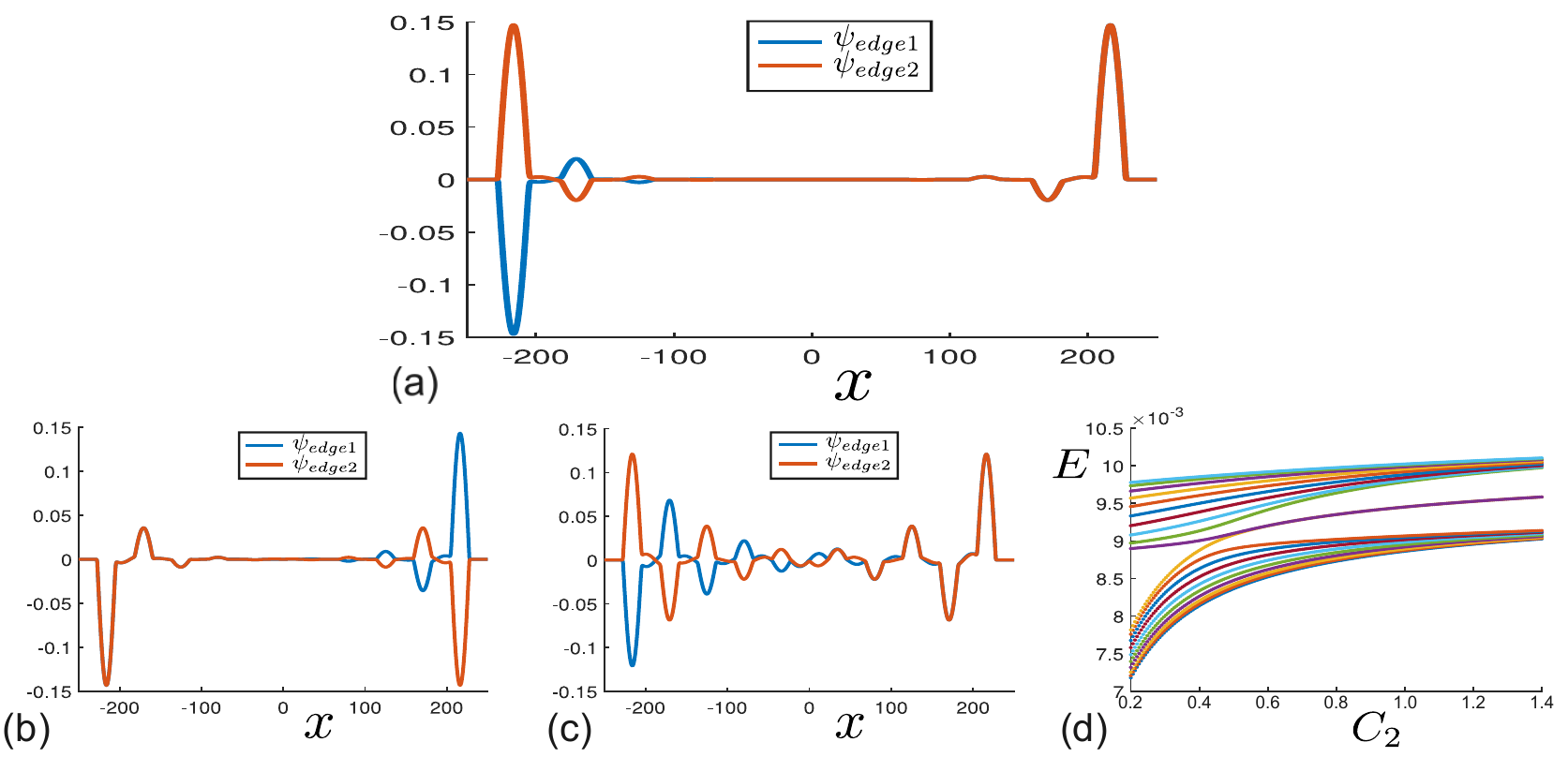}    \caption{llustrates the system's analogy to the SSH model. (a) shows highly localized edge states reflecting the result of chiral symmetry in SSH system, with the left edge confined to the first sublattice and the right to the second. Panels (a)–(c) depict the evolution as \(C_2\) (the amplitude of \(V_2\)) decreases, from higher than \(C_1\) (fixed at 0.4)to approaching similar values: (a) \(C_2 = 1.4\), (b) \(C_2 = 1.0\), (c) \(C_2 = 0.6\), with $ d = 0.004 $ for all cases. The energy diagram in (d) shows edge states in the mid-gap after band closure, calculated from the eigenvalue solution using finite differences.}
    	\label{fig:boundarystates}
    \end{figure*}
     The system represents the physical picture corresponding to the model considered in the previous section, as may be noted for a number of reasons. The system is set up in an ultra-cold environment, leading to low-momentum states. The alternating potentials (illustrated in Figure 5 and defined in Equation 18) generate SSH-like staggered quantum tunneling amplitudes (Equation 1), leading to the formation of two sub-lattices(1 and 2). The high step potentials at the boundaries mimic the open boundaries of a finite SSH system.\\
	 The properties arising as a result of this configuration confirm that of SSH-like Hamiltonian, as is demonstrated in Figure 6. It hosts edge states, i.e., highly localized states at the edges, with each side (right and left) residing on a particular sublattice, i.e., the left edge resides on sublattice A, and the right edge resides on sub-lattice B (particularly refer to Figure 6(a)). As the staggered quantum tunneling amplitude becomes more uniform, the edge states progressively delocalize, as shown in Figures 6(a) to 6(c). These edge states have energies within the band gap, and as the tunneling becomes equal, the edge states merge with the bulk, causing the energy band to close at the transition point, as illustrated in Figure 6(d). The topology associated with this kind of alternating potential has been experimentally confirmed in \cite{atala2013direct} and is theoretically studied in \cite{smith2020topological}. \\
	
	\begin{figure*}
		\centering
		\includegraphics[width=1\linewidth]{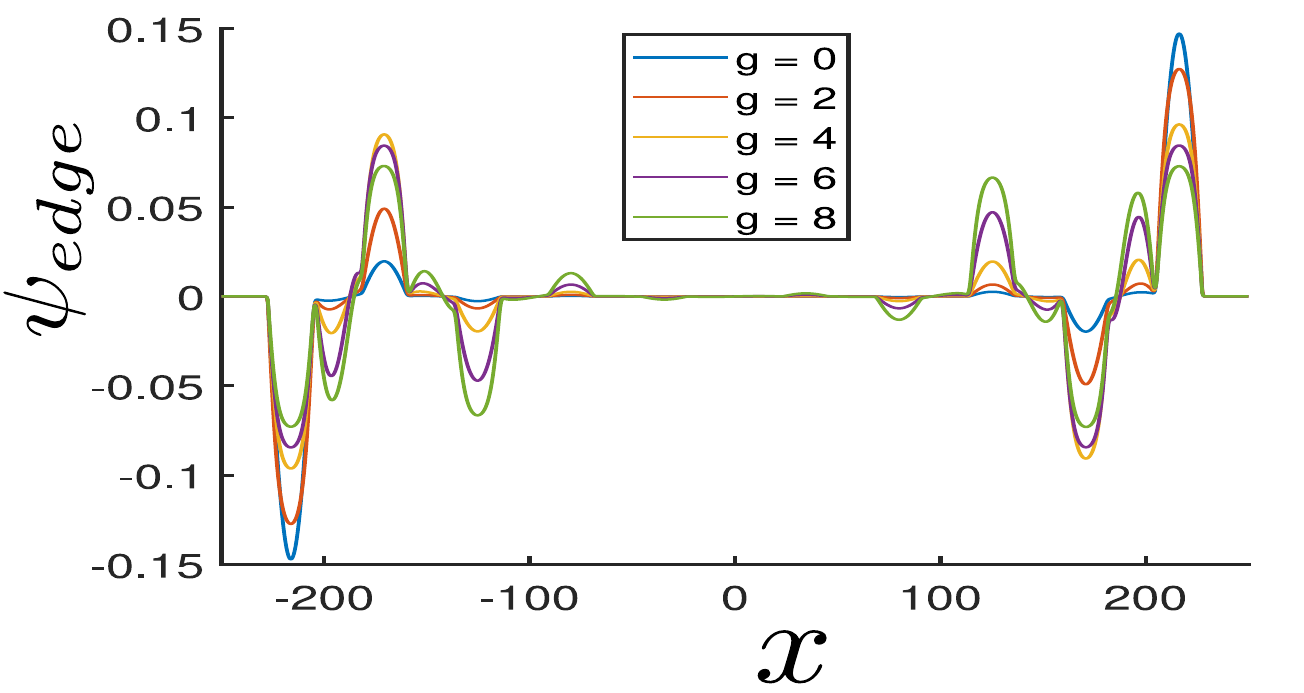}
		\caption{Effect of interactions on boundary wavefunctions: Interactions reduce the wavefunction amplitude at the boundary and enable occupation of sublattices previously restricted by the SSH-like nature of the system, with occupation increasing as interaction strength grows(denoted by $ g $).}
		\label{fig:boundaryint}
	\end{figure*}
	
	Now, we proceed to investigate the boundary states with the interactions present. As the boundary states are not the lowest eigenstates in the system, we use the ITPUSH algorithm developed in Section 3 to study them. The $\psi_{0}$ and $E_{0}$ required for this algorithm are obtained by solving the finite difference eigenvalue problem when no interaction is present. Then, we slowly ramp up the interaction and study the boundary wavefunction for each case. Figure 7 shows the wavefunctions corresponding to different interaction strengths for the configuration corresponding to the wavefunction in Figure 6(a). These results closely align with those obtained through the continuum mechanics approach discussed in the previous subsection.
	It may be noted from Figure 7 that the interactions have the effect of flattening the edge-state, i.e., the amplitude of the wavefunction at the boundary reduces, and it spreads more into the bulk. Another very important effect is populating the sub-lattice that was previously unoccupied for the case of a non-interacting SSH type system, i.e., the $ B $ sub-lattice for the case of the left edge and the $ A $ sub-lattice for the case of the right edge. The occupation of these previously empty sub-lattices gradually increases with an increase in interaction. These effects align with the findings in Section IV A, where the occupation of sublattice B increases with interaction strength, leading to a reduction in occupation on sublattice A. This reduction, however, is not very apparent in Figure 4 due to the difference in magnitude between the two. 
    The momentum-space wavefunctions corresponding to the boundary states calculated in this section as shown in Figure 6 and Figure 7, is confined to region in $ k $ space which is significantly less than $ \pi $. This allows a valid comparison between the overall envelope of these boundary states and those obtained in the previous section using the envelope function approximation. 
    It is worth noting that with this form of potential, which creates staggered tunneling amplitudes, depending on how the boundary is constructed, there may exist other boundary states that do not exhibit the topological nature of the SSH-type Hamiltonian. These states, along with the effects of interactions on them, are examined in Appendix A.
	

	\section{summmary and conclusions}
	Building on the imaginary time propagator method, we developed an algorithm to target specific eigenstates within a spectrum. This algorithm requires an initial guess for both the trial wavefunction and energy, then evolves the wavefunction in imaginary time under the influence of a squared Hamiltonian to converge on the desired state. We applied this approach to investigate the effects of interactions on boundary states of Bosons in a SSH lattice within ultracold settings. 
	
	First, using the continuum approximation, we derived an effective Hamiltonian from the functional minimization of a mean-field energy functional. The algorithm was then used to identify boundary states in the energy spectrum, which included both positive and negative values. Initially, the method was applied to non-interacting Bosons, confirming well-known states across varying levels of boundary sharpness. Next, the boundary states were examined under different interaction strengths, revealing that interactions tend to increase population on the B sublattice while reducing the population on the A sublattice.\\
	We then explored a physical model with analogous characteristics representing the theoretical continuum model: ultracold Bosons trapped in a box containing periodic potential structure with alternating amplitudes. First, the system's properties were examined in the non-interacting regime, drawing an analogy to the SSH model regarding edge states and energy structure. Using the developed algorithm, we then investigated the effect of incorporating interactions into the system on the boundary states. A notable impact of incorporating interactions into the system is that it allows occupation of sub-lattices previously constrained by the SSH nature of the potentials, with this occupation increasing as the interaction strength grows. This behavior aligns with the continuum theory of the problem. Additionally, the interaction causes a reduction in the edge state's amplitude at the boundary, while simultaneously promoting its spread into the bulk.

	\section*{ACKNOWLEDGMENTS}
    A. Ghosh would like to express his gratitude to the University of Melbourne and the Indian Institute of Technology Kharagpur for providing a conducive research environment. A. Ghosh is thankful for the encouragement and support from his group members at both institutes.
    
    \appendix
    \section{Additional Types of Boundary States and the Impact of Interactions}
    Here, we will discuss the impact of interaction on boundary states, which are different from that of the ones discussed in the main paper, which reflected the topological character of the edge states of the SSH type Hamiltonian, i.e., a mid-band positioned with selective sublattice occupation. 
    The potentials considered here have roughly the same form, i.e., consist of alternating potentials(as in Equation 17), except the boundary potentials have the same height as that of one of the inner periodic potentials($ V_{2} $), as illustrated in Figure 8. 
    \begin{figure}[h]
    	\centering
    	\includegraphics[width=1.0\linewidth]{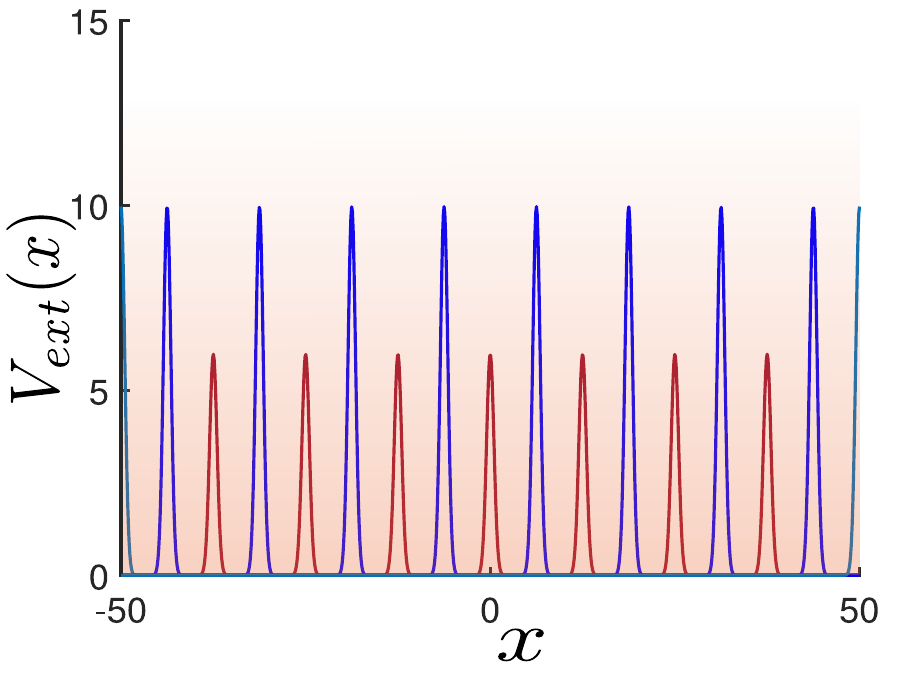}
    	\caption{This figure illustrates the schematic of the system. The deep wells with alternating amplitudes ($V_1$ and $V_2$) result in alternating hopping amplitudes, with the boundary potential set to $V_1$.}
    	\label{fig:externalpotential}
    \end{figure}
	The mathematical form is,
	\begin{equation}\label{key}
		V_{ext}=\sum_{n=2}^{N_{g}-1}[V_{1}[1-mod(n,2)]+V_{2}[mod(n)]]+V_{1}[\delta_{1,n}+\delta_{N_{g},n}],
	\end{equation}
    where $N_g$ represents the total number of Gaussians, including the boundary terms.
	For the above choice of potentials, the energies of these states matches with that of the bulk-states for the case when $ V_{1} = V_{2} $, and as a result gradually delocalized in the bulk as $ V_{1} $, approaches $ V_{2} $.  Figure 9, demonstrates this, (a) to (d) represents the gradual de-localization of these states in to the bulk and (e) represents the corresponding energies, with the distinct higher energies corresponds to these energy states  (rescaled to maintain the boundary state energy at zero).
	\begin{figure}[H]
		\centering
		\includegraphics[width=0.7\linewidth]{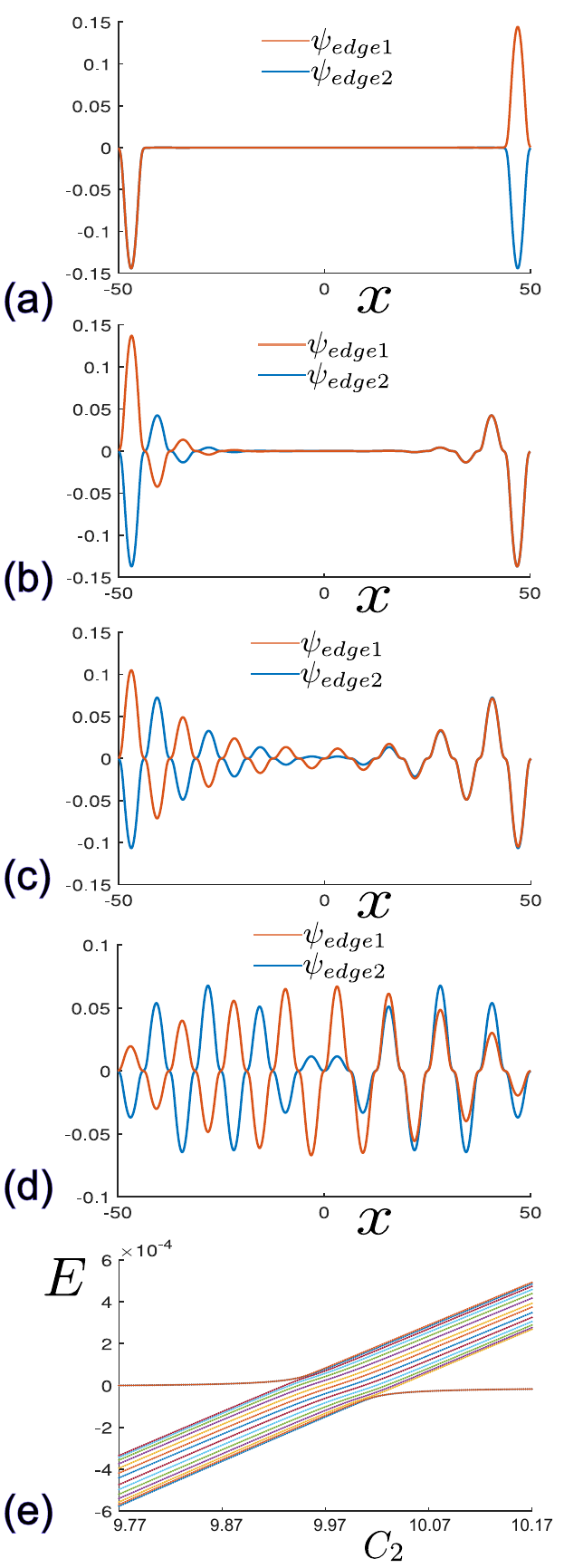}
		\caption{The figure shows boundary states' behavior as staggered nature of the potentials are reduced. Panels (a)–(d) illustrate the progression with increasing \(C_2\) (from 0 to 9.97, with \(C_1\) fixed at 9.97, and $ d $ fixed at $ 0.005 $), resulting in uniform hopping. Panel (e) displays the energy diagram from the finite difference method, indicating that boundary states remain isolated for \(C_2 < C_1\) and a new boundary state appears when \(C_2 = C_1\). The lack of sublattice specificity and energy locations differs from the SSH-type Hamiltonian's topological features. }
		\label{fig:boundarystates}
	\end{figure}
	However, these states may have energies which is much above and below than that of the bulk-states depending upon the form of boundaries, preventing their merging with the bulk in such cases.

Turning to the effect of interactions on these boundary states, we will once again utilize the ITPUSH algorithm introduced in Section 3 of the main text to study them since these boundary states corresponds to the highest eigenstates within the considered energy band. As before, the initial wavefunction $\psi_0$ and energy $E_0$ required for this algorithm are obtained by solving the finite difference eigenvalue problem without interaction. The interaction is then gradually introduced in the Gross-Pitaevskii framework(Equation 16), and the boundary wavefunctions are analyzed. Figure 10 presents the wavefunctions corresponding to various interaction strengths.It can be observed that interaction causes the wavefunctions to spread, distributing in a manner that minimizes the increased potential energy resulting from the interaction.
\\

	\begin{figure}[H]
		\centering
		\includegraphics[width=1\linewidth]{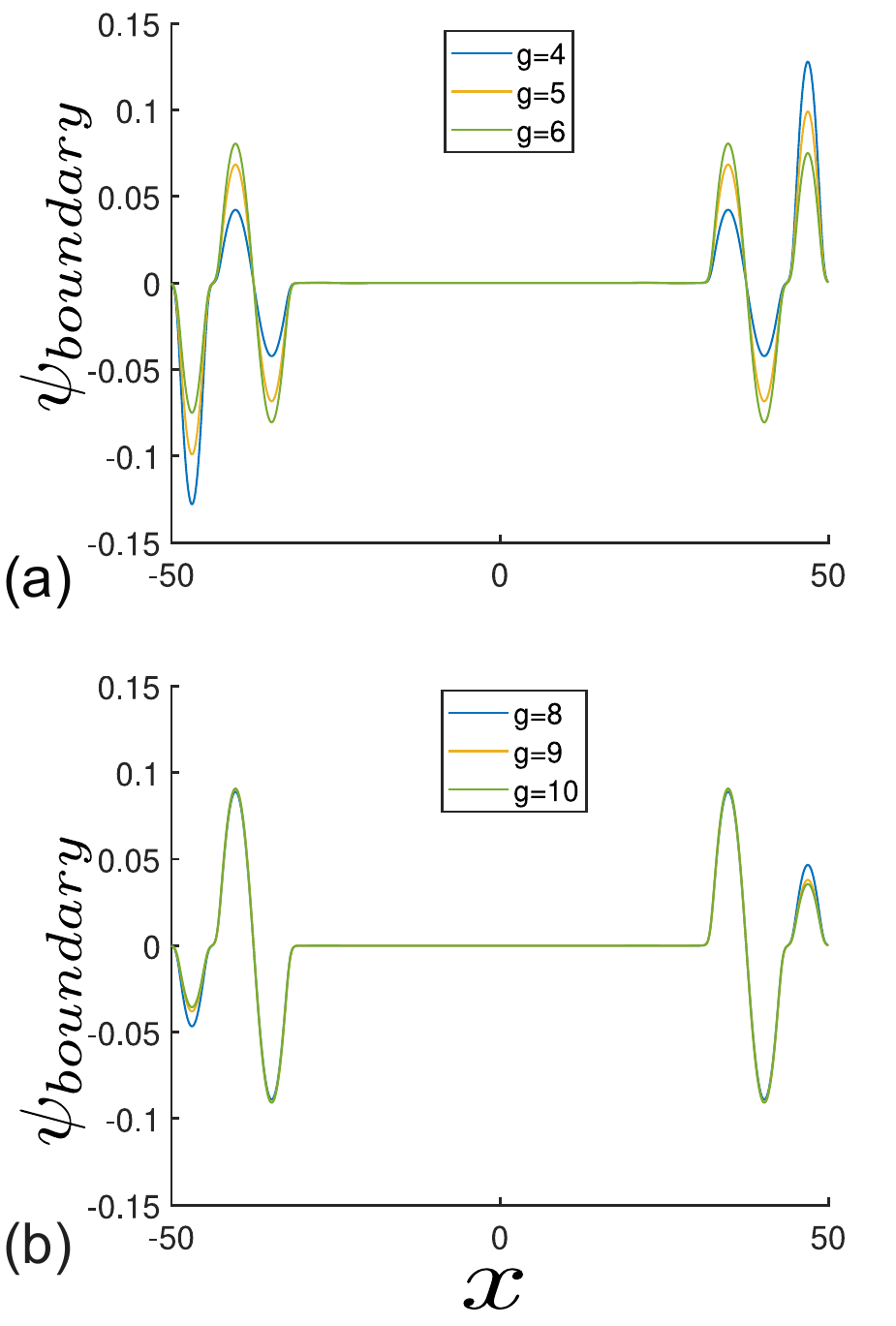}
		\caption{The impact of interactions on the boundary wavefunctions. (a) shows the boundary wavefunctions for interactions \( g = \{4, 5, 6\} \), and (b) for \( g = \{8, 9, 10\} \). A gradual flattening of the boundary wavefunctions is observed with increasing interactions.}
		\label{fig:boundaryint}
	\end{figure}
	\bibliographystyle{unsrtnat}
	\bibliography{References}

\end{document}